\documentclass[final,5p,times,twocolumn]{elsarticle}

\usepackage[english]{babel}
\usepackage[hidelinks]{hyperref}
\makeatletter\AtBeginDocument{\let\@elt\relax}\makeatother

\usepackage{booktabs}
\usepackage{longtable}
\usepackage{tabularx}
\usepackage{graphicx}
\usepackage{amsmath}
\usepackage{float}
\usepackage[english,capitalise]{cleveref}
\usepackage{mathtools}
\usepackage{bbm}

\usepackage[utf8]{inputenc}

\usepackage{siunitx}
\usepackage[dvipsnames,usenames]{xcolor}
\definecolor{green}{rgb}{0.2,0.6,0.2}

\usepackage{physics}
\usepackage{csquotes}
\usepackage{multirow}
\usepackage{ifthen}
\usepackage{ulem}
\usepackage{orcidlink}

\usepackage{relsize} 

\renewcommand{\v}[1]{\ensuremath{\boldsymbol{#1}}}

\newcommand{\K}[1]{\ensuremath{\left(#1\right)}}

\newcommand{\ci}{\mathrm{i}}

\makeindex

\renewcommand{\vec}[1]{\v{#1}}

\newcommand{\jmi}{j_{\mathrm{max},\mathrm{int}}}
\newcommand{\jmf}{j_{\mathrm{max},\mathrm{free}}}
\newcommand{\lmi}{l_{\mathrm{max},\mathrm{int}}}
\newcommand{\lmf}{l_{\mathrm{max},\mathrm{free}}}

\newcommand{\jmis}{j_{\mathrm{mi}}}
\newcommand{\jmfs}{j_{\mathrm{mf}}}

\newcommand{\docdate}{19.05.2025}

\usepackage{etoolbox}
\patchcmd{\MaketitleBox}{\footnotesize\itshape\elsaddress\par\vskip36pt}{{\footnotesize\itshape\elsaddress}\par\vskip15pt {\docdate} \par\vskip20pt}{}{}

\patchcmd{\pprintMaketitle}{\footnotesize\itshape\elsaddress\par\vskip36pt}{{\footnotesize\itshape\elsaddress}\par\vskip15pt {\docdate} \par\vskip20pt}{}{}

\begin{document}

\begin{frontmatter}

\title{Nucleon-nucleon correlation functions from different interactions in comparison}

\author[infn]{Matthias Göbel \orcidlink{0000-0002-7232-0033}}
\ead{matthias.goebel@pi.infn.it}

\author[infn]{Alejandro Kievsky \orcidlink{0000-0003-4855-6326}}
\ead{alejandro.kievsky@pi.infn.it}

\affiliation[infn]{organization={Istituto Nazionale di Fisica Nucleare, Sezione di Pisa},
    addressline={Largo Pontecorvo 3}, 
    city={56127 Pisa},
    country={Italy}
}


\date{\docdate}

\begin{abstract}

Correlation functions as they can be observed in heavy-ion collisions using the femtoscopy technique
are a powerful tool to study the interaction among different baryons or mesons.
Specifically, the multi-nucleon correlation functions have been under intense experimental and theoretical investigation in the recent years.
Due to the interest of using this observable as an input in the construction of potentials
between hadrons we revisit the nucleon-nucleon correlation function and calculate it
using different nuclear interactions at high precision. Since the nucleon-nucleon potential is determined to reproduce the two-nucleon scattering data, we would like to critically evaluate the amount of this information captured by the correlation function.
We study the dependence of the correlations on the nuclear force giving detailed insights into the calculations, in particular the convergence behavior in the partial waves. The coupling between the different partial-wave channels is taken into account and the relevance of this effect is quantified.
To make contact with precedent studies the results based on the Argonne V18 interaction are presented.
Then we consider also the Norfolk NV2-IIa and NV2-IIb chiral EFT interactions.
The analysis of the differences between the correlations of the various interactions shows that for momenta
between 0 and 500~MeV there are variations of up to 5.9~\% for the \(nn\) system, of up to 1.8~\% for the \(np\) system, and of 1.4~\% for the \(pp\) system.

\end{abstract}

\end{frontmatter}


\section{Introduction}
\label{sec:intro}

The correlation between multiple hadrons or other particles measured in heavy-ion collision is an interesting observable,
which is experimentally well feasible at accelerator facilities like CERN \cite{Wiedemann:1999qn,Heinz:1999rw,Lisa:2005dd,Fabbietti:2020bfg}.
Let us illustrate the correlation function at the example of two emitted particles.
Being a function of the relative momentum between the two particles it is the fraction of particle pairs having this momentum
in relation to the overall number of particles.
This correlation function is mainly influenced by three physical conditions:
the emission process, the particle statistics (fermionic or bosonic), and the interaction among the particles after emission, which falls into
the category of final-state interactions\footnote{
    Of course, there is also interaction between the measured particles and potential other particles which were part of the initial
    heavy-ion collision.
    However, the influence of these interactions is usually negligible.
}.
Initially, the interest focused on the correlation functions as a way to gain knowledge about the emission process,
which can be modeled by a source function.
Now, that the emission process is better understood, there is also growing interest in the correlation function as a manner
to study the consequent final-state interaction.
In the case of two nucleons it is given by the nucleon-nucleon interaction.
Especially in the field of the hyper-nuclear interaction there is much interest in the correlation functions, since, in contrast to the nucleon-nucleon case, for the nucleon-\(\Lambda\) system there is only very limited scattering data available.
Therefore, nucleon-\(\Lambda\) correlation functions could be an important way to constrain the nucleon-\(\Lambda\) interaction.
Studies along this line just started \cite{Mihaylov:2023ahn}.
Testing our understanding of the nuclear interaction via femtoscopy has been also done in the three-body sector.
Measurements of $pd$, $ppp$ and $pp\Lambda$ correlations are now available \cite{ALICE:2023bny,ALICE:2022boj} and, at the same time,
calculations of the $pd$, \(nn\Lambda\), \(pp\Lambda\), \(nnn\), and \(ppp\) system have been performed \cite{Viviani:2023kxw,Kievsky:2023maf,Garrido:2024pwi}.
Thereby the three-body scattering had to be solved. This was achieved with the hyperspherical harmonics method and
the hyperspherical adiabatic basis.

In this paper, we would like to return to the two-body case, to study from a theoretical point of view the nucleon-nucleon correlation function as a method to constrain
the nucleon-nucleon interaction.
Although, in contrast to the hypernuclear sector there exists many scattering data for the nucleon-nucleon system, correlation function data could still provide a supplemental way to constrain the interaction in detail.
And, more important, we want to understand which information is captured by the correlation function.
This is especially relevant to potential efforts to constrain the nucleon-\(\Lambda\) interaction from this quantity.
We shed light onto this by quantifying the sensitivity of the correlation function on different nuclear interactions.
The more recent developments of the nuclear interaction are the introduction of realistic interactions and later the state-of-the-art chiral EFT interactions.
The realistic interactions are based on meson-theory, but are ultimately phenomenological fits.
They provide a good description of the scattering data, important interactions are AV18 \cite{Wiringa:1994wb} and CD-Bonn \cite{Machleidt:1987hj,Machleidt:2000ge}.
In the chiral EFT interactions the interaction terms are given by an effective description of QCD with nucleons, pions, and in some cases
also Deltas as degrees of freedom.
In the construction of the nuclear potential pion-exchanges are already iterated so that a pure \(NN\) interaction is constructed.
Nowadays, many chiral interactions are available. Among other things, they differ in the fit, in the chiral order, the inclusion of \(\Delta\) excitations, the regulators, and the space in which they are formulated in (momentum space or coordinate space).
For reviews on this topic see Refs. \cite{Epelbaum:2008ga,Hammer:2019poc,Epelbaum:2019kcf}.
In this paper, we will limit the study to local interactions. We use the chiral Norfolk interactions \cite{Piarulli:2014bda,Piarulli:2016vel} and additionally consider the AV18 interaction, which was already used to study the \(pp\) correlation function. It should be noticed that the $pp$ correlation function has a low energy peak mostly determined by the $^1S_0$ state. We study in detail the formation of the peak showing that it strongly depends on the $pp$ scattering length and effective range.


\section{The two-body correlation function}
\label{sec:fd}

A formal definition of the correlation at a nucleon-nucleon relative momentum of \(k\) can be given in terms of the probability ratio
\begin{equation}
    C({\bf k}_1,{\bf k}_2)=\frac{{\cal P}({\bf k}_1,{\bf k}_2)}{{\cal P}({\bf k}_1){\cal P}({\bf k}_2)}
\end{equation}
with ${\cal P}({\bf k}_1,{\bf k}_2)$ the probability of having simultaneously two nucleons with momenta ${\bf k}_1,{\bf k}_2$ and ${\cal P}({\bf k})$ the single probability. Following Ref. \cite{Mrowczynski:2020ugu}, this ratio can be described as a process in which the particles are emitted with the subsequent interaction of the nucleon pair 
\begin{align}
    C({\bf k}_1,{\bf k}_2) &=\frac{1}{N}\sum_{m_1,m_2}\int d{\bf r}_1 \int d{\bf r}_2 S_1({\bf r}_1) S_1({\bf r}_2) \nonumber \\
    &\qquad \times |\Psi_{m_1,m_2}({\bf k}_1{\bf r}_1,{\bf k}_2{\bf r}_2)|^2
\end{align}
with $\Psi_{m_1,m_2}({\bf k}_1{\bf r}_1,{\bf k}_2{\bf r}_2)$ 
the two-nucleon scattering wave function with spin projections $m_1,m_2$ and the spin weight $N=(2s_1+1)(2s_2+1)=4$.
In the above equation $S_1({\bf r})$ represents the source function for the emission of a nucleon. After integrating out the center of mass dependence and rewriting the nucleon wave funcion as  
$\Psi_{m_1,m_2}({\bf k}_1{\bf r}_1,{\bf k}_2{\bf r}_2)= e^{-\ci {\bf R}\cdot{\bf P}} \Psi_{m_1,m_2,{\bf k}}({\bf r})$,
with ${\bf R}$ the center of mass coordinate, ${\bf r}$ the relative distance and ${\bf k}=({\bf k}_1-{\bf k}_2)/2$,
the Koonin-Pratt relation is obtained \cite{Koonin:1977fh}
\begin{equation}
    C({\bf k})=\frac{1}{N}\sum_{m_1,m_2}\int d{\bf r} S({\bf r}) |\Psi_{m_1,m_2,{\bf k}}({\bf r})|^2 \, .
\end{equation}
Averaging on the momentum directions, it can be written as
\begin{equation}
    C{(k)} = \frac{1}{N} \sum_{m_1,m_2} \int \dd{\Omega_{\v{k}}} \mel{\Psi_{m_1,m_2,\v{k}}}{ S(\bf r) }{\Psi_{m_1,m_2,\v{k}}} \,.
\end{equation}
Thereby, we have expressed it as the expectation value of the source function
\(S\) evaluated at the position operator $\bf r$.
For the correlation function, the interaction between the nucleons after their creation in the heavy-ion
collision has to be taken into account.
Therefore, the expectation value is not based on a plane-wave state but on the nucleon-nucleon scattering state with \(\v{k}\)
as outgoing momentum denoted as \(\ket{\Psi_{m_1,m_2,\v{k}}}\).

We calculate the correlation function in a partial-wave basis for the two-nucleon system.
The orbital angular momentum is quantized by \(l\) and the overall spin by \(s\).
Together they coupled to the total angular momentum  \(j\) with projection \(m\).
The isospin is specified by \(t\) with projection \(m_t\).
Evaluating the expectation value in a coordinate-space partial-wave basis the correlation function reads
\begin{equation}
    C{\K{k}} 
    = \frac{1}{N} \sum_{l,s} \sum_{j,m} \sum_{t,m_t} \sum_{l'}  \int \dd{r} r^2 S{\K{r}}  
    \left| \Psi_{k;\K{l,s,}j,m,t,m_t}^{(l')}{\K{r}} \right|^2 \, ,
\end{equation}
where from now on we consider a spherically symmetric source.
The partial-wave projected wave function is given by
\begin{equation}
    \Psi_{k;\K{l,s,}j,m,t,m_t}^{(l')}{\K{r}} = \braket{r;\K{l',s}j,m,t,m_t}{\Psi_{k;\K{l,s}j,m,t,m_t}} \,.
\end{equation}
Since the \(NN\) scattering is diagonal in all quantum numbers except for \(l\) due to the tensor force, only for
\(l\) we have to distinguish between the ingoing and the outgoing quantum number.
The outgoing is \(l\), while the ingoing is \(l'\).

In practice, we calculate \(\Psi_{k;\K{l,s,}j,m,t,m_t}^{(l')}{\K{r}}\) only for the lowest partial waves, in which the radial wave functions are distorted by the \(NN\) interaction.
For the higher partial waves we use the free (or Coulomb) wave functions up to some truncation.
The criterion for the truncation of the partial wave series will be that \(C{(k \to \infty)} \to 1\) is fulfilled to a good degree.
One can either realize the truncation in the \(l\) quantum number or in the \(j\) quantum number.
For the most purposes, we will truncate in \(j\).
When truncating the interacting wave functions at \(\jmi=\jmis\) and the free wave functions at \(\jmf=\jmfs\),
the expression for the correlation function reads
\begin{align}
    C{\K{k}} 
    &= \frac{1}{N} \sum_{0 \leq j \leq \jmis} \sum_{l,s,t} \sum_{l'}  w_{j,t} \int \dd{r} r^2 S{\K{r}}  
    \left| \Psi_{k;\K{l,s,}j,t}^{(l')}{\K{r}} \right|^2 \nonumber \\
    &\quad + \frac{1}{N} \sum_{\jmis < j \leq \jmfs} \sum_{l,s,t} w_{j,t} \int \dd{r} r^2 S{\K{r}}  
    \left| \Psi_{\mathrm{free};k;\K{l,s,}j,t}{\K{r}} \right|^2 
\end{align}
Note that hereby we used also the fact that the wave function is usually independent of \(m\) and \(m_t\). Due to evaluating the corresponding sums the weight factor \(w_{j,t} = \K{2j+1} \K{2t+1}\) was introduced.

We calculate the correlation functions for different nucleon-nucleon interaction:
the Argonne V18 interaction \cite{Wiringa:1994wb} as well as the Norfolk I/II interactions \cite{Piarulli:2014bda,Piarulli:2016vel}.
The electromagnetic interactions are in all cases the ones of Ref. \cite{Wiringa:1994wb}.
This means that for the \(pp\) system there are one- and two-photon Coulomb terms, the Darwin-Foldy term, the vacuum polarization as well as the magnetic-moment interaction.
For the \(np\) system, there is Coulomb interaction due the finite charge distribution as well as the magnetic-moment interaction.
In the case of the \(nn\) system, solely the magnetic-moment interaction is taken into account.

In addition to the wave function, another important ingredient in the computation of the correlation function is the source function. Following previous studies, we use a Gaussian parametrization for the single source function. Due to its simplicity, the Gaussian parametrization is commonly used in the description of the correlation function. However other forms with a more complex angular dependence have been studied before \cite{Danielewicz:2005qh,ALICE:2017iga}. After integrating out the center of mass coordinate, one obtains that also the two-nucleon source function is of Gaussian shape
\begin{equation}
    S{(r)} = \K{4\pi \rho^2}^{-3/2} e^{-r^2 / (4\rho^2)} \,
\end{equation}
with the source size \(\rho\).
Indications on the source size $\rho$ can be obtained using the transverse mass $m_T$ scaling with the transverse mass defined as $m_T=(k_T^2+m^2)^{1/2}$, where $k_T$ is the average transverse momentum and $m$ is the average mass of the pair. For proton-proton collisions a typical value is \(\rho=1.249\)~fm \cite{ALICE:2019buq}.

In the following we provide a brief overview on the calculation of the wave function.
In the case that there is no coupling between different channels, the reduced radial wave function
\(u_{p;(l,s)j,t}(r)\) fulfills the radial Schrödinger equation
\begin{equation}
    \K{ \partial_r^2 - l(l+1) r^{-2} - \widetilde{V}_{(l,s)j,t}{(r)} + k^2} u_{k;(l,s)j,t}{(r)} = 0 \,,
\end{equation}
with the rescaled potential \(\widetilde{V}_{(l,s)j,t}{(r)} = 2\mu V_{(l,s)j,t}{(r)}\).
The reduced radial wave function fulfills the boundary condition
\begin{equation}
    u_{k;(l,s)j,t}{(r)} \rightarrow \widetilde{j}_l{(kr)} + e^{\ci \delta_l(k)} \sin{\K{\delta_l{(k)}}} \widetilde{h}^+_l{(kr)} \,.
\end{equation}
Hereby, \(\widetilde{j}_l\) is the Riccati-Bessel function and \(\widetilde{h}_l^+ = \widetilde{n}_l + \ci \widetilde{j}_l\) is the Riccati-Hankel function. The Riccati-Neumann function is given by \(\widetilde{n}_l\).
The relation between the Riccati-Bessel function and the spherical Bessel function \(j_l\) is given by \(\widetilde{j}_l{(z)} = z j_l{(z)}\).

When the Coulomb interaction is present, \(\widetilde{j}_l\) is replaced by the regular Coulomb wave function \(F_l\) and
\(\widetilde{n}_l\) is replaced by the irregular Coulomb wave function \(G_l\).
The relation between the radial wave function in a certain partial wave and the reduced radial wave function is given by
\(\Psi_{p;(l,s)j,t}{(pr)} = 4\pi \ci^l u_{p;(l,s)j,t}{(pr)}/(pr)\).

Due to the presence of the nuclear tensor force there are cases where the interaction
couples channels with the same \(j\) and \(s\) but different \(l\).
Specifically, the \(s=1\) channels with \(l=j-1\) and \(l=j+1\) are coupled.
In this \(s,j,t\) subchannel the potential turns into a 2x2 matrix with non-vanishing off-diagonal entries.
The radial wave function is a two-element vector with the \(l=j-1\) and \(l=j+1\) radial wave functions as entries.
The coupled-channel Schrödinger equation reads
\begin{align}
    \K{\partial_r^2 - \frac{\hat{l}(\hat{l}+1)}{r^2} + k^2} \vec{u}_{k;s,j,t}{(r)} - \widetilde{V}_{s,j,t}(r) \vec{u}_{k;s,j,t}{(r)} = 0 \,,
\end{align}
whereby \(\hat{l}\) is the 2x2 diagonal matrix with the corresponding values of \(l\) on the diagonal.
By solving this equation two times with different initial values and some algebra, one can reconstruct the two fundamental solutions
\(\vec{u}^{(l_-)}_{k;s,j,t}{(r)}\) and \(\vec{u}^{(l_+)}_{k;s,j,t}{(r)}\), which differ in the incoming partial wave.
The boundary condition for \(\vec{u}^{(l)}_{k;s,j,t}{(r)}\) is given by
\begin{align}
    \vec{u}^{(l)}_{k;s,j,t}{(r)} =
    \begin{pmatrix} u^{(l)}_{k;(l_-,s),j,t}{(r)} \\ u^{(l)}_{k;(l_+,s),j,t}{(r)} \end{pmatrix}
    \rightarrow
    \begin{pmatrix} 
        \delta_{l,l_-} \widetilde{j}_l{(kr)} + T^{(l,l_-)}_{k;s,j,t} \widetilde{h}^+_{l_-}{(kr)} \\
        \delta_{l,l_+} \widetilde{j}_l{(kr)} + T^{(l,l_+)}_{k;s,j,t} \widetilde{h}^+_{l_+}{(kr)}
    \end{pmatrix} \,,
\end{align}
with \(T^{(l,l')}_{k;s,j,t}\) being the t matrix in that channel at that momentum \(k\).
The uncoupled and coupled ordinary differential equations (ODEs) are solved using an implementation of the Runge-Kutta-Fehlberg method,
a method of order 4 combined with an error estimation of order 5.
This allows for adaptive adjustments of the step width.


\section{Results}
\label{sec:results}

\subsection{The composition of the \(pp\) correlation function and universality}
\label{ssec:struc_pp_cf}

\begin{figure}[H]
    \centering
    \includegraphics[width=0.45\textwidth]{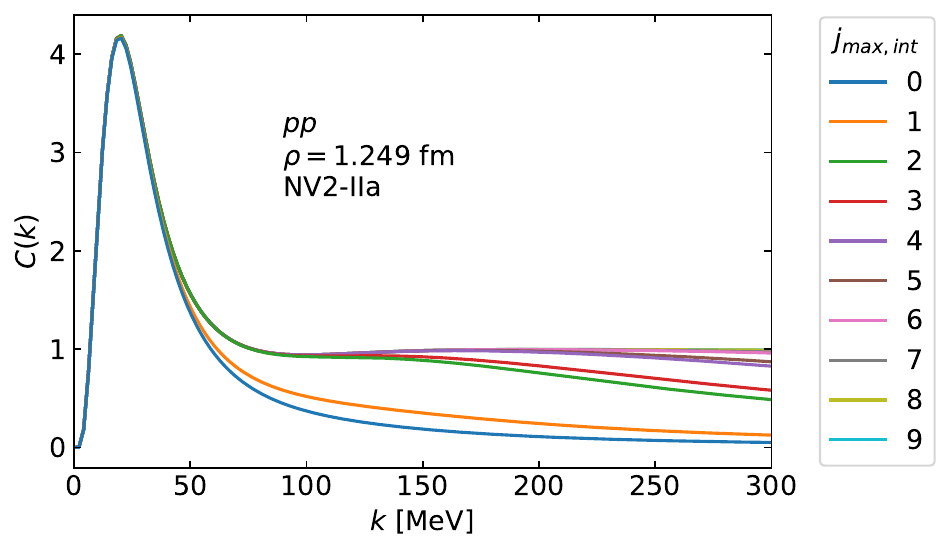}
    \caption{The \(pp\) correlation functions obtained with different \(\jmi\) are shown.
    They are solely based on interacting components (\(\jmf = \jmi\)). NV2-IIa is used.
    }
    \label{fig:cf_pp_pw_cmps_cumul}
\end{figure}

We start with illustrating the decomposition in terms of components of definite spin \(j\) of typical \(NN\) correlation functions at the example
of the \(pp\) correlation function obtained with the NV2-IIa interaction and a source radius
of 1.249~fm.
The result is shown in \cref{fig:cf_pp_pw_cmps_cumul}.
The upper panel shows correlation functions differing in \(\jmi\). \(\jmf=\jmi\) was used so that
all correlation functions are only based on interacting components, i.e., components obtained by solving the interacting Schrödinger equation.

\begin{figure}[hbt]
    \centering
    \includegraphics[width=0.45\textwidth]{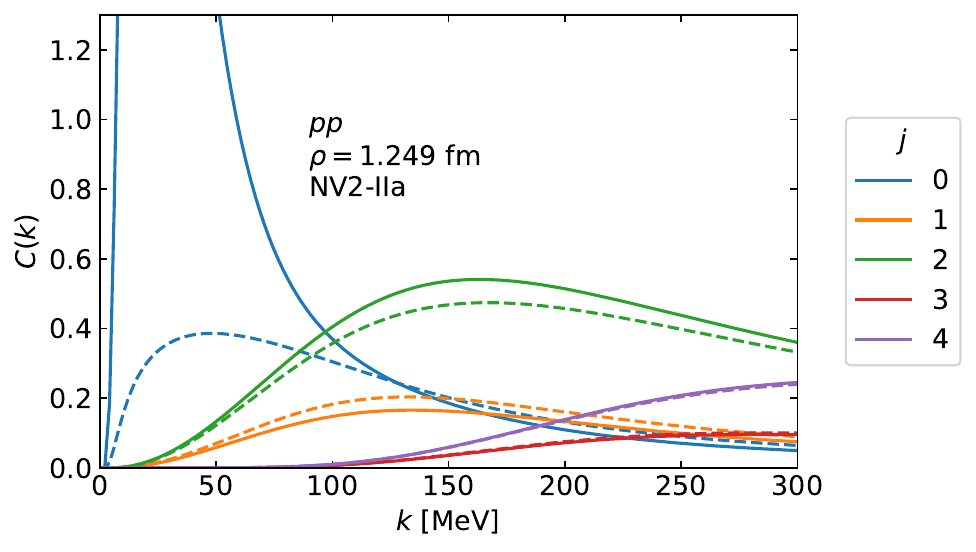}
    \caption{The upper panel shows the \(pp\) correlation functions obtained with different \(\jmi\).
    They are solely based on interacting components (\(\jmf = \jmi\)).
    The lower panel shows a smaller collection of components corresponding to a single \(j\). components.
    The solid lines are interacting contributions, while the dashed lines show the free contributions.}
    \label{fig:cf_pp_pw_cmps}
\end{figure}

We observe that for \(k \leq 50\)~MeV the correlation function is dominated by the \(j=0\) component of the wave function and for \(k \leq 100\)~MeV using \(\jmi=3\) seems to be sufficient.
Moreover, we are interested in the difference between the free and the interacting results on the level of the single \(j\) components.
\Cref{fig:cf_pp_pw_cmps} shows the interacting components (solid lines) in comparison with the corresponding free components (no interaction taken into account, dashed lines).

This shows that in absolute numbers this difference is largest for the \(j=0\) contribution. Due to the centrifugal barrier
it gets smaller with increasing \(j\).
For the \(pp\) (as well as for the \(nn\) system) the \(j=0\) component has two subcomponents: \(l=0, s=0, j=0\) and \(l=1, s=1, j=0\).
It might be especially 
interesting to see how large is the \(s\)-wave contribution to the formation of the low-energy peak.

\begin{figure}[H]
    \centering
    \includegraphics[width=0.4\textwidth]{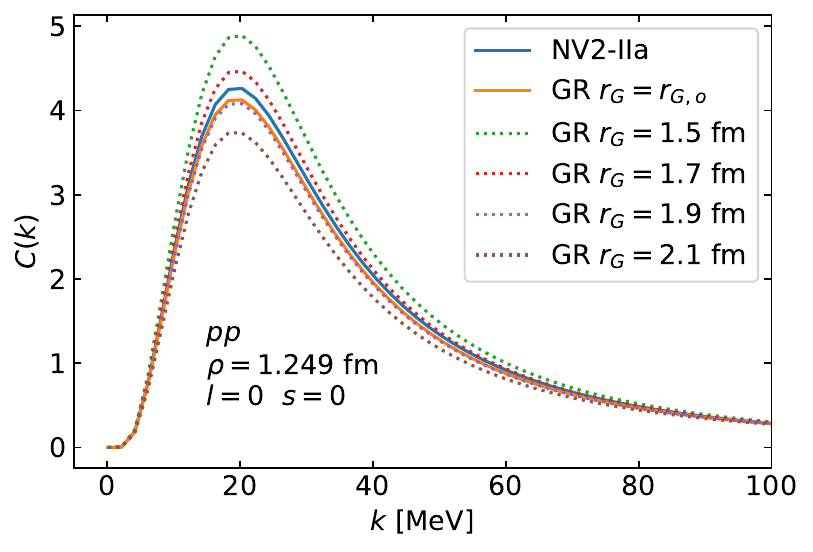}
    \caption{
    The \(s\)-wave \(pp\) correlation from NV2-IIa in comparison with Gaussian representations (GR).
    The Gaussian representations differ in \(r_G\) values.
    The first one (\(r_G = r_o\)) was obtained by fitting \(V_0\) and $r_G$
    to the \(s\)-wave scattering length $a_0$ and effective range of NV2-IIa.
    In the other cases \(r_G\) were fixed at the specified values,
    while \(V_0\) was tuned to reproduce \(a_0\).}
    \label{fig:cf_pp_pw_cmp_gr}
\end{figure}

To quantify this we looked at the fraction of the correlation functions obtained with a certain \(\lmi=\lmf\) in relation to a reference calculation based on \(\lmi=10\) and \(\lmf=35\).
Up to \(k \approx 25\)~MeV, the \(s\)-wave contribution constitutes more than 98~\% of the overall correlation function.
Using \(\lmi=2\) one can almost go as high as \(k=100\)~MeV while maintaining the same accuracy level. And all that without even using additional free components. In practice one would employ these, but the purpose of this specific analysis is to highlight the importance of certain contributions.
We conclude that the relevance of the \(s\)-wave component in the low-momentum region and especially to the peak is high.
A plot depicting the fractions of the specific correlation functions can be found in the supplementary material of this letter.
To investigate further the sensitivity of the peak to the low-energy scattering parameters we can use universal concepts \cite{Kievsky:2021ghz}. To this end we construct a low-energy Gaussian representation for the \(s\)-wave interaction.
\Cref{fig:cf_pp_pw_cmp_gr} compares the \(s\)-wave $pp$ correlation function component using the NV2-IIa and different Gaussian representations defined as
\begin{equation}
V_G(r)=V_0 e^{-(r/r_G)^2} + e^2 r^{-1}    
\end{equation}
where, for different choices of the Gaussian range $r_G$, the Gaussian strength $V_0$ is determined to reproduce the $pp$ scattering length of $-7.8\,$fm. In this way the Gaussian potential captures the short-range interaction coming from the strong force and for the electromagnetic force as discussed in Ref. \cite{Tumino:2023yfz}.

For the comparison of the \(s\)-wave correlation function with the Gaussian representations, we observe a good description overall. It is especially good if not only the Gaussian potential's strength parameter but also its range is used as fitting parameter for matching the scattering length and the effective range.
If only the strength parameter is fitted to the scattering length, the results are still qualitatively correct.

The optimal representation deviates from the NV2-IIa correlation function by less than 5~\% in the depicted region, whereby the deviation at the peak position is less than 3~\%.
The fact that a simplified interaction, the Gaussian representation, which captures essential features of the phase shifts yields in this momentum region a good description shows that the correlation function is largely constrained by the phase shift.
Approximations to single-partial-wave components of the correlation function can also be obtained from effective-range expansion parameters, see, e.g., Ref. \cite{Albaladejo:2025kuv}.

\subsection{The \(pp\), \(np\) and \(nn\) correlation functions}
\label{ssec:cfs}

\begin{figure*} 
    \centering
    \includegraphics[width=0.97\textwidth]{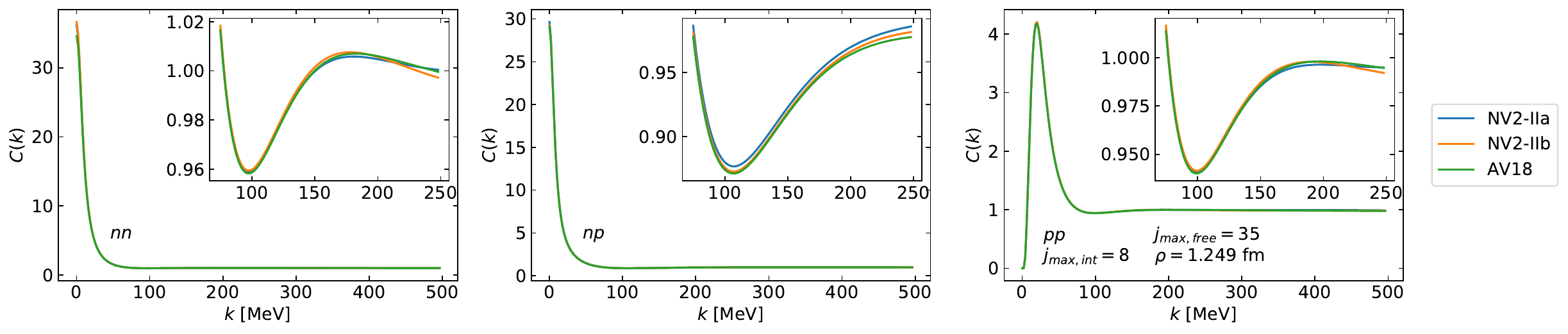}
    \caption{The left panel shows the \(nn\) correlation function, the middle panel shows the
             \(np\) correlation function, and the right panel shows the \(pp\) correlation functions.
             The different line colors correspond to the different nucleon-nucleon interactions in use.}
    \label{fig:cfs_all}
\end{figure*}

We now look at the \(nn\), \(np\) and \(pp\) correlation functions themselves.
They are shown for different interactions in \cref{fig:cfs_all}.
The truncation parameters are \(\jmi=8\) and \(\jmf=35\).
The left panel shows the \(nn\) correlation function, the middle panel the \(np\) one, and the
right panel shows the \(pp\) correlation function.

The plots show that asymptotically all three correlation functions go to one as expected.
Moreover, one can see that for \(k \to 0\) the \(nn\) and \(np\) correlation function take finite values
greater than one, while the \(pp\) correlation function goes to zero.
This because at small \(k\) the overall interaction, i.e., the nuclear and the electromagnetic interaction, is
attractive in the \(nn\) and the \(np\) system, while it is repulsive the \(pp\) system
due to the Coulomb force.
In addition to the peak and tail structure, all three correlation functions feature also a dip around \(k = 100\)~MeV.
While the dip position is for the \(nn\) and \(pp\) system close to around 97~MeV and 99~MeV, it is for the \(np\) system with about 107~MeV considerably larger.
The minimum value of the dip is smaller for the \(np\) system than for the \(nn\) or \(pp\) system.
While it is about 0.96 for the \(nn\) system and circa 0.94 for the \(pp\) system, it amounts to circa 0.87 for the \(np\) system. The dip is a result of an interference between \(s\)- and \(p\)-wave function components. A good description of it requires interacting components at least up to \(j=2\). 
Moreover, one can see that results are qualitatively the same for all considered interactions (AV18, NV2-IIa, NV2-IIb).

We observe that there are some small differences between the results obtained with different interactions.
In the region between \(k=100\)~MeV and \(k=250\)~MeV this effect is strongest in the \(np\) system.
In the dip region NV2-IIb and AV18 are similar, while NV2-IIa is a bit away.
Around 250~MeV, the deviations seem to be more equal with NV2-IIb in the middle.
That the deviations are more visible for the \(np\) system than for the \(nn\) system is a bit surprising, given
that the interaction in the \(np\) system is already well constrained by scattering data.
However, one has to take into account that the deviations are in both cases, also for the \(np\) system, very small.
Moreover, this is only a statement about the region between 100~MeV and 250~MeV.
Recognizing deviations at low momenta is more difficult due to the large slopes observed there.
We will study the sensitivity on the interaction in more detail in \cref{ssec:int_sens} by looking at deviations
in percent.
Instead of the NV2-II interactions, one can also calculate correlation functions for the NV2-I interactions.
The difference of these potentials is the fit region used to determine the low-energy constants.
While NV2-II uses a maximum \(T_{\mathrm{lab}}\) of 200~MeV, NV2-I corresponds to a maximum lab energy of 125~MeV.
In terms of the momentum \(k\) these energies correspond to circa 307~MeV and 242~MeV.
Because of the smaller fitting region, in the case of the NV2-I we see for the \(nn\) and the \(pp\) system stronger deviations
already starting around \(k=150\)~MeV.
In other words: In those momentum regions where the different interactions yield similar phase shifts also the produced correlations are similar.
In addition to our earlier study with the Gaussian representation, this is another indication that the Koonin-Pratt formula defines a quantity strongly constrained by phase shifts and thereby might be a good approximation to a strict observable quantity.
Thereby correlation function data are a useful supplement to scattering data, especially in cases where the latter are rare.
Plots of the correlation functions for the NV2-I interactions can be found in the appendix \cref{ap:cfs_nv2_i}.
A detailed discussion of the uncertainties of our results can be found in the supplementary material.
The uncertainty sources are the truncation in partial waves as well as the numerical uncertainties of solving the ODEs for the
scattering wave functions.
We estimate that the relative uncertainties are below 0.2~\%.

\subsection{Influence of the source radius}
\label{ssec:rho}

\begin{figure*} 
    \centering
    \includegraphics[width=0.97\textwidth]{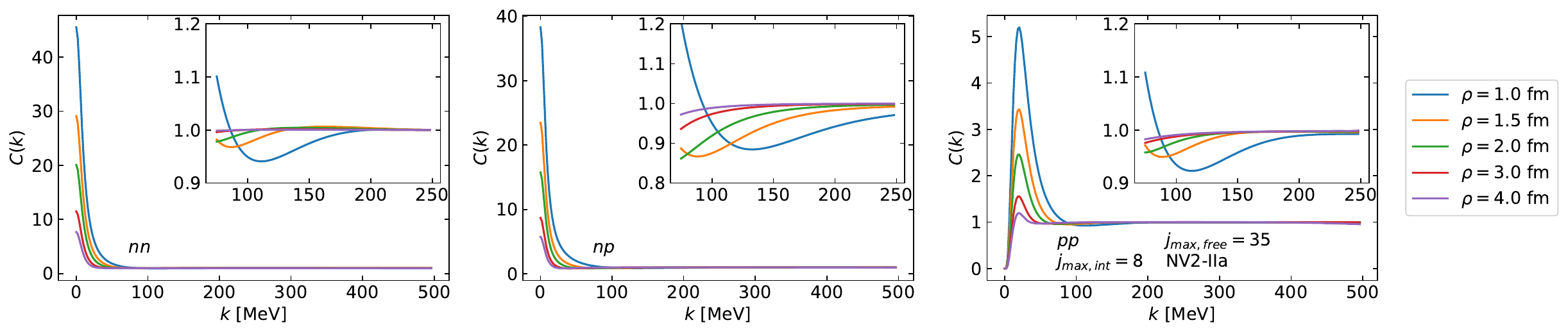}
    \caption{The correlations functions are depicted for the different source radii \(\rho = 1.0\), \(1.5\), \(2.0\), \(3.0\), \(4.0\)~fm.
    The left panel shows the \(nn\) correlation function, the middle panel shows the \(np\) correlation function, and the \(pp\)
    one is depicted in the right panel.
    All correlation functions were obtained with the interaction NV2-IIa.}
    \label{fig:cfs_all_rho}
\end{figure*}

We now study the correlation functions obtained with different source radii for the Gaussian source in use.
The correlation functions obtained with NV2-IIa and different source radii are depicted in \cref{fig:cfs_all_rho}.

We observe that the smaller the source radius the higher is the peak.
This is true for the \(pp\) system where the peak is position at a finite \(k\), but also for the \(nn\) and \(np\) system,
where the peak is at zero momentum.
Moreover, there is also a trend for the minimum (dip) position.
For all systems with decreasing source radii the dip position goes to higher momenta.
For \(nn\) it moves from about 73~MeV for \(\rho=1.0\)~fm to about 111~MeV for \(\rho=2.0\)~fm.
For \(pp\) it changes from about 75~MeV to 113~MeV.
In the case of \(np\), the dip position moves from 65~MeV to 133~MeV.
In the case of the \(nn\) and the \(pp\) system smaller source radii also correlate with a deeper dip.
The change between \(\rho=1.0\)~fm and \(\rho=2.0\)~fm is from 0.94 to 0.98 for \(nn\) and from 0.92 to 0.96 for \(pp\),
while it is from 0.89 to 0.85 for \(np\).
I.e., \(np\) shows the opposite behavior. Correlation functions with different source radii can be obtained from different heavy ion collisions. Whereas the used radius of \(1.249\)~fm refers to \(pp\) collisions at \(\sqrt{s}=13\)~TeV,
larger radii appear for example in Pb-Pb collisions \cite{ALICE:2015hvw}.

\subsection{Convergence in the included channels}
\label{ssec:ch_cnvg}

\begin{figure}[H]
    \centering
    \includegraphics[width=0.47\textwidth]{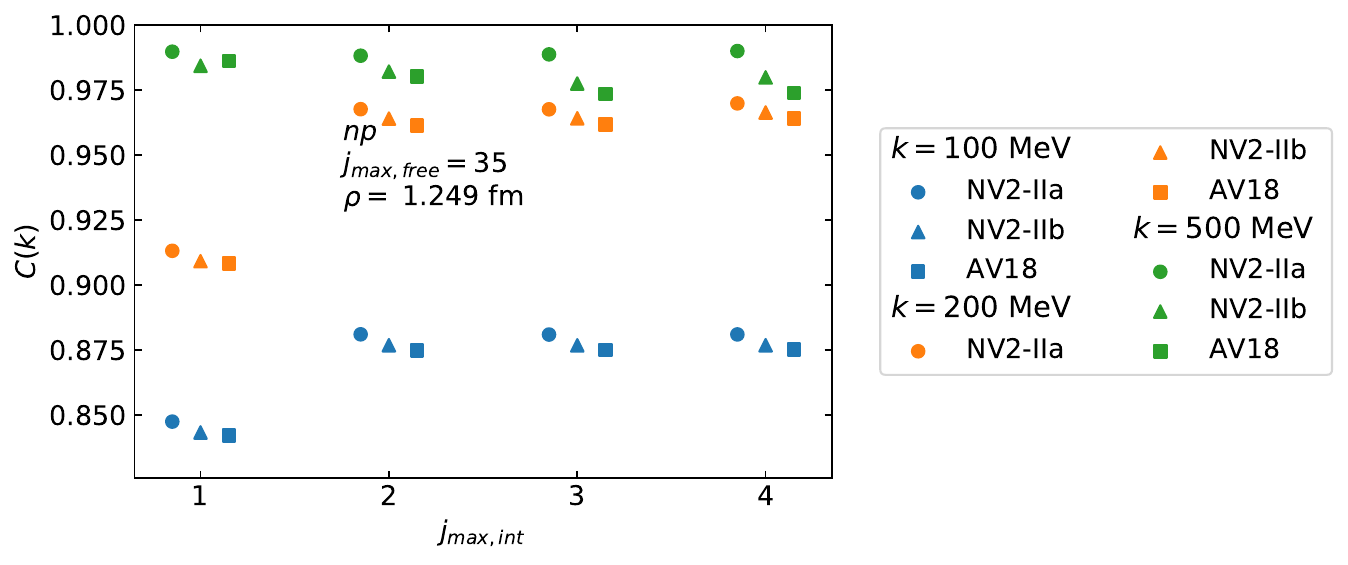}
    \caption{The values of the \(np\) correlation function as function of the \(j\) of the highest interacting partial wave given
    by \(\jmi\). The color of the symbol indicates the momentum \(k\), while the shape indicates the employed interaction.}
    \label{fig:cf_np_jmi_pw_cvg}
\end{figure}

Another interesting aspect of the study of the correlation functions is their convergence in the partial waves.
This is additionally interesting in regards to understanding the deviations between different interactions.
One can check if the sensitivity to the interaction in a certain momentum region can be attributed to a specific partial wave.
\Cref{fig:cf_np_jmi_pw_cvg} shows the \(np\) correlation function values at certain \(k\) for different interactions as function of
the truncation in the included interacting partial-waves given by \(\jmi\).

It can be seen that there is good convergence in \(\jmi\) for quite some different momenta ranging between 100 and 500~MeV.
For relative momenta of 100 and 200~MeV approximate convergence is already reached at \(\jmi = 2\).
At 500~MeV at least the interactions in \(j=3\) should be also included, i.e., using at minimum \(\jmi = 3\) is recommended.
By looking at how the differences between the differently-shaped symbol behave we can check where the sensitivity on the interaction enters.
For \(k=200\)~MeV, we see that some of the sensitivity enters in \(j=2\).
For \(k=500\)~MeV, there is more sensitivity. It mainly stems from \(j=2\) and from \(j=3\).

\begin{figure}[H]
    \centering
    \includegraphics[width=0.45\textwidth]{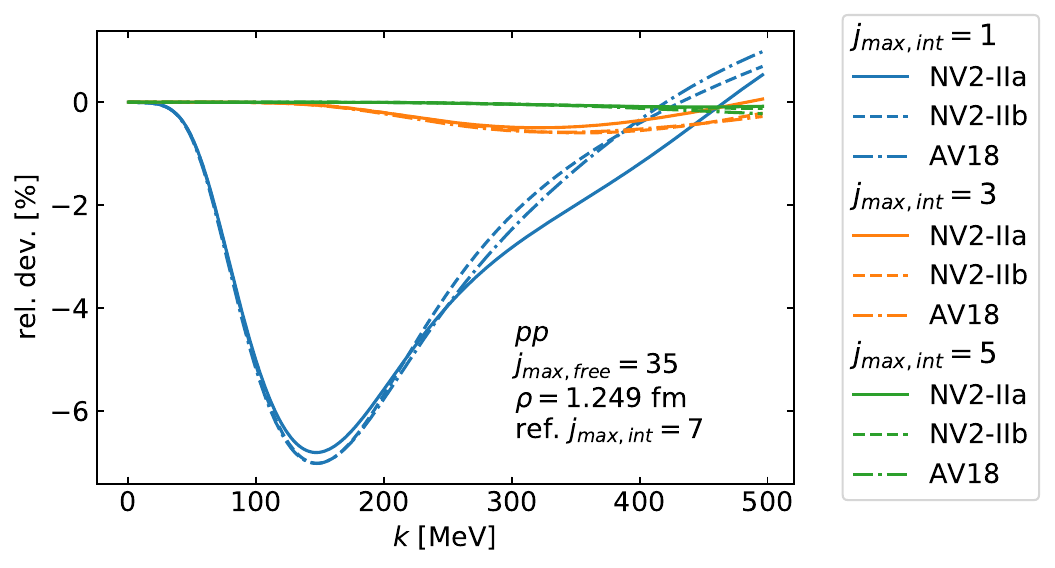}
    \caption{The relative deviation of \(pp\) correlation functions obtained at different \(\jmi\) from the corresponding correlation function
    obtained with \(\jmi=7\) is shown.
    The line styles encode the underlying interaction.}
    \label{fig:cf_pp_jmi_rel_diff}
\end{figure}

The discussed plot is useful for investigating the contributions from different partial waves, especially in regards to sensitivity.
However, one might also want to study the convergence in \(\jmi\) over the complete momentum range of interest.
For this purpose, \cref{fig:cf_pp_jmi_rel_diff} shows the relative deviations in percent of \(pp\) correlation functions obtained at different \(\jmi\) 
from the correlation function obtained at \(\jmi=7\).
Written as a formula the depicted quantity is given by \(100\K{C{(k;\jmi)/C{(k;\jmi=7)-1}}}\).

We conclude that the convergence behavior is similar for the different interactions and good over the complete momentum range
under consideration.
We did the same analysis also for the \(nn\) and \(np\) systems.
A plot depicting this quantity for all three systems is contained in the supplementary material.
Considering all three systems, we conclude that the results obtained with \(\jmi = 1\) deviate up to about 7.2~\%, whereby the maximum deviation is reached between 100 and 200~MeV.
At \(\jmi=3\) the deviation from the \(\jmi = 7\) results is already below 0.7~\%, and for \(\jmi=5\) any deviation is hardly visible.
It is below 0.25~\%.
As a consequence results obtained with \(\jmi\) around 7 are very well converged.

\subsection{Influence of the coupling of channels}
\label{ssec:cplg}

Sometimes, simplified calculations of correlation functions are performed where the coupling between different partial-wave channels is neglected.
In the case of the nuclear interaction the tensor force couples the \(s=1\) channels with \(l=j-1\) and \(l'=j+1\).
We investigate the influence of the couplings by calculating the relative deviation of the correlation function without coupling
from the one with coupling included.
For the \(nn\) and the \(pp\) system we observe that the absolute value of the relative deviation doesn't get much larger than 1~\%.
It is reached in the region between 100 and 200~MeV.    
In the case of the \(np\) system, the effect is much larger.
Depending on the interaction in use, the deviation can reach up to about 30~\%. 
Another difference is here the position of the maximum. It is located below 100~MeV.
That the coupling effects for the \(np\) system are much larger than for the \(nn\) and \(pp\) system can be understood
in terms of the partial-wave channels which are coupled.
For the \(nn\) and \(pp\) system the first channels which are coupled, are \(^3P_2\) and \(^3F_2\).
These channels are also present for the \(np\) system, but there is also coupling between the channels \(^3S_1\) and \(^3D_1\).
The latter are in the \(nn\) and \(pp\) system not present due to antisymmetrization (\(\K{-1}^{l+s+t} = -1\) has to hold).
Therefore, for \(np\) coupling occurs already in lower partial waves and as the interactions are stronger in the lower partial
waves, the effects of the coupling tends to be stronger.
A plot of the coupling effects is contained in the supplementary material.

\subsection{Sensitivity on the interaction}
\label{ssec:int_sens}

\begin{figure*}
    \centering
    \includegraphics[width=0.97\textwidth]{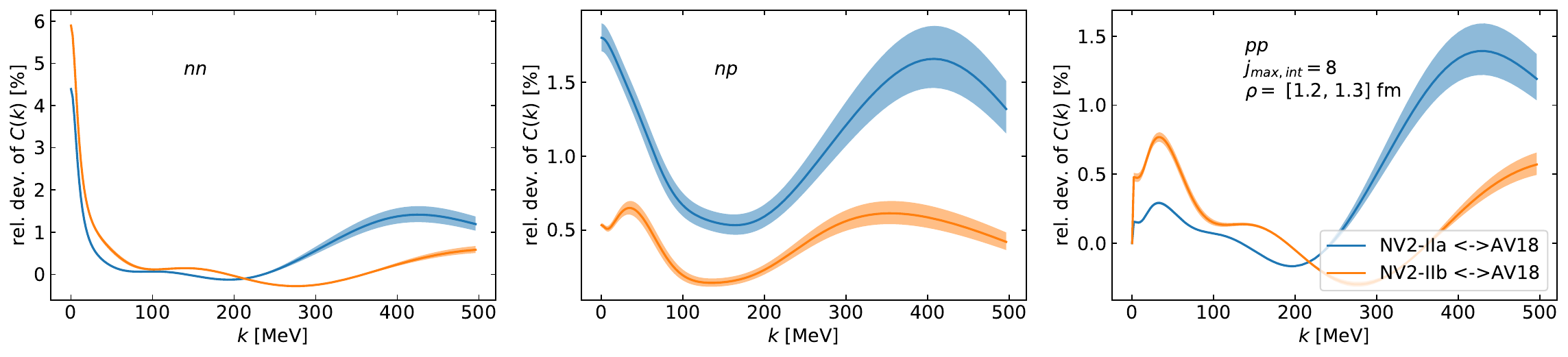}
    \caption{The relative deviations of the NV2-IIa and NV2-IIb correlation functions from the AV18 correlation functions are shown in percent in order to study the sensitivity on the interaction.
    The panels differ in the physical system: \(nn\), \(np\), and \(pp\) (from left to right).
    The uncertainty bands correspond to a variation of \(\rho\) between 1.2~fm and 1.3~fm.}
    \label{fig:cf_all_rel_diff}
\end{figure*}

Finally, we want to study the sensitivity of correlation functions on the \(NN\) interaction in more detail.
For strange systems such as the \(p\Lambda\) system there are only a few scattering data and recently correlation functions
are used as a means to constrain the interactions.
For the \(NN\) system, mainly for the \(np\) and the \(pp\) systems, there are much more scattering data. Nevertheless,
it could also be useful to use data from correlation functions in the fitting process of the low-energy constants of chiral interactions.
For that reason, we want to study how sensitive the correlation function is on the difference between a few popular interactions,
namely AV18, NV2-IIa, and NV2-IIb.
The relative deviation of the NV2-IIa and the NV2-IIb correlation functions from the AV18 correlation function is shown
in \cref{fig:cf_all_rel_diff} for the different systems.
As formula the quantity reads \(100\K{C{\K{k;\mathrm{NV2-IIx}}}/C{\K{k;\mathrm{AV18}}}-1}\) (with \(\mathrm{x} \in \{\mathrm{a},\mathrm{b}\}\)).
An uncertainty band corresponding to a variation of the source radius between 1.2~fm and 1.3~fm is also shown.
This variation is of the order of typical experimental uncertainties for the source radius.
In the femtoscopy experiment reported in Ref. \cite{ALICE:2019buq} in \(pp\) collisions a source radius \(\rho=1.249_{-0.029}^{+0.032}\)~fm was determined.

We observe that for the \(nn\) and the \(np\) system the largest sensitivity in the region from \(k=0\)~MeV to \(k=500\)~MeV is located at zero momentum. In contrast to that, the \(pp\) interaction displays the largest deviation around \(k=430\)~MeV.
In the case of the \(np\) and \(pp\) systems NV2-IIa displays the largest deviation from AV18. For the \(nn\) system it is NV2-IIb.
Quantitatively, we find that the deviation of NV2-IIb from AV18 for the neutrons peaks at 5.9~\%.
For the \(np\) system, the deviation of NV2-IIa goes up to 1.8~\%.
In the case of the protons, the correlation function of NV2-IIa differs by up to 1.4~\%.
Note that these results are very well converged. The uncertainties from the truncation in the partial waves and from solving the ODEs
for obtaining the scattering wave function are estimated to be smaller than 0.1~\%.
This is not an absolute percentage, but a relative measure to the results displayed in \cref{fig:cf_all_rel_diff}.
More information can be found in the supplementary material.

\begin{figure}[H]
    \centering
    \includegraphics[width=0.45\textwidth]{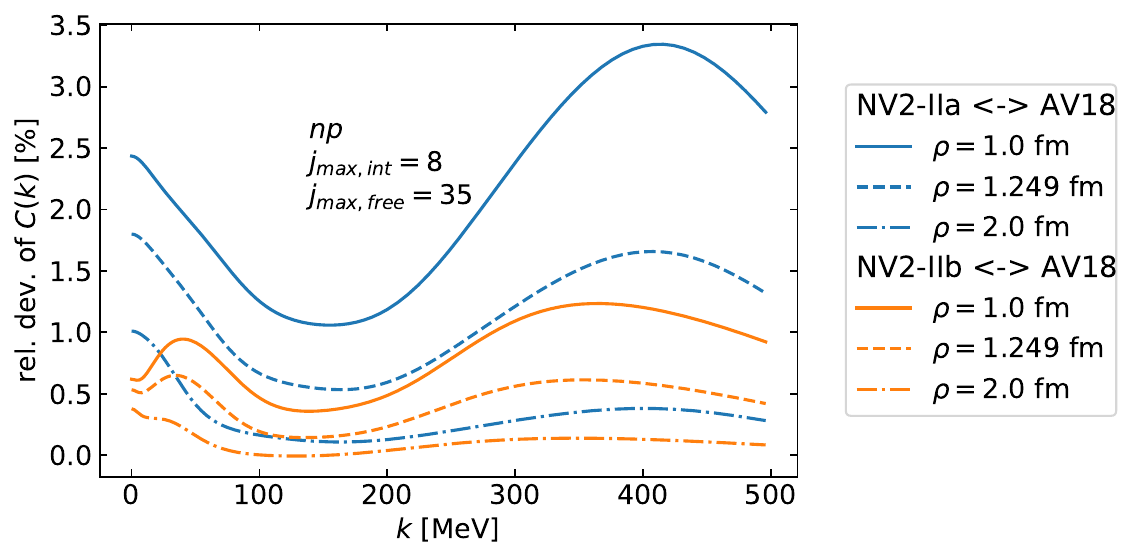}
    \caption{The relative deviation of the NV2-IIa and NV2-IIb correlation functions from the AV18 one
    are depicted for the \(np\) system for different source radii.
    The different source radii correspond to different line styles.}
    \label{fig:cf_np_mult_rhos}
\end{figure}

While the uncertainty bands in the figure based on the variation of the source radius between 1.2~fm and 1.3~fm indicate that uncertainties in this order do not have a large influence on the sensitivity, it is also interesting to investigate how larger changes
in the source radius would influence the sensitivity.
\Cref{fig:cf_np_mult_rhos} displays the sensitivities for the \(np\) system for the source radii of 1.0~fm, 1.249~fm, and 2.0~fm.

It shows that the larger the source radius the smaller is the sensitivity of the correlation function on the differences between the investigated interaction.
While for low momenta near 0, the damping in the sensitivity related to going from 1.0~fm to 2.0~fm is between 1 and 3, it is between \(k=300\)~MeV and \(k=500\)~MeV more around 9. 
At high momenta, an increase of the source radius would affect the sensitivity much more.


\section{Conclusion \& Outlook}
\label{sec:conclusion}

The nucleon-nucleon correlation functions based on different interactions,
specifically AV18 and the Norfolk chiral interactions, have been investigated in detail.
For that purpose we solved the two-body scattering Schrödinger equation to obtain the scattering wave function
and calculated on this basis the correlation function.
By including interactions in the partial waves up to \(j=8\) and by including the non-interacting contributions
up to \(j=35\), we obtain highly accurate results. Coupling due to the tensor force is taken into account.
We illustrated the composition of the nucleon-nucleon correlation function at the example of the \(pp\) system.
The low-momentum peak is almost entirely caused by the \(s\)-wave contribution showing a pronounced sensitivity to the low-energy scattering parameters, the scattering length and effective range. At higher momenta contributions from other partial waves become more important. 
Next, we studied the correlation functions of the different two-nucleon systems.
While the \(nn\) and \(np\) correlation function peak at zero momentum, the \(pp\) correlation function peaks around
\(k=20\)~MeV. This is the effect of the Coulomb repulsion.
Asymptotically, all correlation functions go to one.
A characteristic property is the a minimum around 100~MeV.
While for the \(nn\) and \(pp\) systems it is located slightly below 100~MeV, for the \(np\) system it is at circa 107~MeV. The minimum is produced by an interplay between \(s\)- and \(p\)-waves. Accordingly, to properly describe the minimum at least the interaction has to be incorporated up to \(j=2\).

We studied also the effect of the radius of the Gaussian source, it was varied between 1.0~fm and 4.0~fm.
Across all isospin channels, lower source radii yielded higher peak values and the shift of the minimum position to higher momenta.
This position is quite sensitive to the source radius.
Changing the source radius from 1.0~fm to 2.0~fm moves the minimum from approx. 65~MeV to 133~MeV. 
The convergence in the number of interacting partial waves and the effect of the coupling between the different partial-wave channels were investigated in detail.
While for the \(nn\) and the \(pp\) system the influence of the coupling is of the order of 1~\%, depending on the interaction, it reaches almost 30~\% for the \(np\) system.

Moreover, the sensitivity of the correlations on the interaction has been studied.
We used as a measure the deviations in percent of the NV2-IIa and NV2-IIb correlation functions from the AV18 one.
In the \(nn\) case, the maximum deviation is about 5.9~\% at zero momentum mostly due to differences in the value of the \(nn\) scattering length.
Also for the \(np\) system the maximum is reached at zero momentum. It amounts to 1.8~\%.
In contrast to that, the two protons have the peak sensitivity with 1.4~\% around \(k=430\)~MeV.
Recently, the behavior of the correlation function given by the Koonin-Pratt formula under unitary transformations has been studied and discussed in the context of the observability of the correlations \cite{Epelbaum:2025aan}. 
We presented multiple instances where similar phase shifts lead to similar correlation functions.
Therefore, it is likely that the Koonin-Pratt definition is a
good approximation to a strict observable and, under certain limitations, is almost invariant under unitary transformations.
More studies along this line are in progress.
In conclusion we have analyzed in detail the nucleon-nucleon correlation functions showing its sensitivity on the interaction.
This analysis can be seen as a step further in the possibility of using experimentally determined correlations as a supplement to phase shift data in the process of determining interaction's low-energy constants.

\section*{Acknowledgments}

We acknowledge helpful discussions with L. Girlanda, L. E. Marcucci, and M. Viviani.

\appendix


\section{Correlation functions obtained with NV2-Ia and NV2-Ib interactions}
\label{ap:cfs_nv2_i}

\begin{figure}[H]
    \centering
    \includegraphics[width=0.45\textwidth]{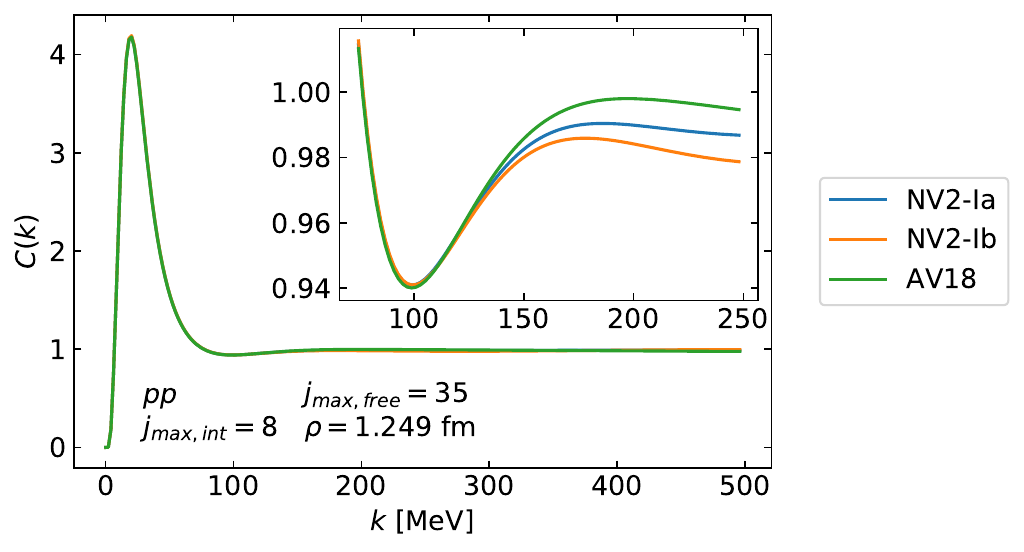}
    \caption{\(pp\) correlation functions obtained with different interactions are shown.}
    \label{fig:cfs_pp_g1}
\end{figure}

In \cref{fig:cfs_all}, we have presented correlation functions obtained with AV18 as well as NV2-IIa and NV-IIb.
We complement this here by showing results for NV2-Ia and NV2-Ib for
the \(pp\) system in \cref{fig:cfs_pp_g1}.

One can see that at higher momenta, starting at around \(k=150\)~MeV, for the \(pp\) system, there is much more difference between the correlation functions from different interactions.
The same is true for the \(nn\) system.
The deviations of the NV2-Ia results from the AV18 results are about 0.01, and the ones of NV2-Ib from AV18 results are about 0.02.
One can understand this difference in terms of the different fitting regions.
While the NV2-I interactions have been fitted to scattering data up to \(T_{\mathrm{lab}}=125\)~MeV, the NV2-II ones have been fitted
up to \(T_{\mathrm{lab}}=200\)~MeV.
In terms of the momentum \(k\) that corresponds approximately to 242~MeV and 307~MeV, respectively.
AV18 was fitted up to \(T_{\mathrm{lab}}=350\)~MeV.

\bibliographystyle{elsarticle-num-names} 
\bibliography{nn_cf_paper.bbl}

\end{document}